# *Investigating the use of publicly available natural videos to learn Dynamic MR image reconstruction*

Dr. Olivier Jaubert[1], Mr. Michele Pascale[1], Dr. Javier Montalt-Tordera[1], Mr. Julius Akesson[1,2], Dr. Ruta Virsinskaite[3], Dr. Daniel Knight[1,3], Pr. Simon Arridge[4], Dr. Jennifer Steeden[1], Pr. Vivek Muthurangu[1]

[1.] UCL Centre for Translational Cardiovascular Imaging, University College London, London. WC1N 1EH. United Kingdom

[2.] Clinical Physiology, Department of Clinical Sciences Lund, Lund University, Lund. Sweden.

[3.] Department of Cardiology, Royal Free London NHS Foundation Trust, London. NW3 2QG. United Kingdom.

[4.] Department of Computer Science, University College London, London. WC1E 6BT. United Kingdom

Corresponding author:

| | |
|---|---|
| Name | Vivek Muthurangu |
| Department | Centre for Translational Cardiovascular Imaging |
| Institute | University College London |
| Address | 30 Guilford St, London WC1N 1EH |
| E-mail | v.muthurangu@ucl.ac.uk |

*Grant Support*: This work was supported by UK Research and Innovation (MR/S032290/1), Heart Research UK (RG2661/17/20) and the British Heart Foundation (NH/18/1/33511, PG/17/6/32797).

*Article Type:* Technical Developments

*Word Count:* 2000

*Acknowledgments*: This work used the public dataset Inter4K and open-source framework TensorFlowMRI (v0.22.0).

*Data Availability Statement:* The source code for training and testing our framework using natural videos and corresponding pre-trained networks for dynamic MRI reconstruction are provided online (https://github.com/olivier-jaubert/Image_Reconstruction_Inter4k.git).

*Summary Statement:*

A pipeline was developed to learn dynamic MR image reconstruction from publicly available natural videos. The pipeline was tested on real-time dynamic MR applications providing similar subjective image quality to conventionally learnt reconstructions while requiring no MR specific data.

*Key Points:*

- A pipeline was developed to transform natural videos into dynamic multi-coil k-space data enabling supervised learning of dynamic multi-coil image reconstruction methods including VarNet, 3D UNet and FastDVDNet.
- In prospective experiments three image orientations were collected in 10 healthy subjects (SAX, 4CH and Speech), the cardiac and natural video deep learning reconstructions were ranked similarly (and higher than CS) and presented no statistical differences in SNR and Edge Sharpness for most conditions.
- Advantages of using natural videos for training include large numbers of high quality, high spatio-temporal resolution data readily available and easily shareable data, code, and models.

# Abstract

*Purpose:* To develop and assess a deep learning (DL) pipeline to learn dynamic MR image reconstruction from publicly available natural videos (Inter4K).

*Materials and Methods:* Learning was performed for a range of DL architectures (VarNet, 3D UNet, FastDVDNet) and corresponding sampling patterns (Cartesian, radial, spiral) either from true multi-coil cardiac MR data (N=692) or from pseudo-MR data simulated from Inter4K natural videos (N=692). Real-time undersampled dynamic MR images were reconstructed using DL networks trained with cardiac data and natural videos, and compressed sensing (CS). Differences were assessed in simulations (N=104 datasets) in terms of MSE, PSNR, and SSIM and prospectively for cardiac (short axis, four chambers, N=20) and speech (N=10) data in terms of subjective image quality ranking, SNR and Edge sharpness. Friedman Chi Square tests with post-hoc Nemenyi analysis were performed to assess statistical significance.

*Results:* For all simulation metrics, DL networks trained with cardiac data outperformed DL networks trained with natural videos, which outperformed CS ($p<0.05$). However, in prospective experiments DL reconstructions using both training datasets were ranked similarly (and higher than CS) and presented no statistical differences in SNR and Edge Sharpness for most conditions. Additionally, high SSIM was measured between the DL methods with cardiac data and natural videos (SSIM>0.85).

*Conclusion:* The developed pipeline enabled learning dynamic MR reconstruction from natural videos preserving DL reconstruction advantages such as high quality fast and ultra-fast reconstructions while overcoming some limitations (data scarcity or sharing). The natural video dataset, code and pre-trained networks are made readily available on github.

Abbreviations: MR = Magnetic Resonance, CS = Compressed sensing, DL = deep learning, SAX = short axis, 4CH = four chamber, SSIM = Structural Similarity Index Measure, SNR = Signal to Noise Ratio, PSNR = Peak Signal to Noise Ratio, MSE = Mean Squared Error.

# Introduction

Real-time magnetic resonance (MR) imaging allows evaluation of dynamic changes without relying on periodic motion. Real-time MR is used to assess the heart, evaluate speech and measure bowel motility(1). However, real-time imaging often requires significant data undersampling to ensure adequate spatio-temporal resolution. Therefore, it is usually combined with advanced reconstruction techniques to produce artifact-free images, with the previous state-of-the-art being Compressed Sensing (CS) (2).

Unfortunately, there are drawbacks to CS including computationally intensive, time-consuming reconstructions and unnatural looking images. Recently, it has been shown that Deep Learning (DL) can outperform CS in terms of reconstruction time and image quality(3). However, DL reconstruction is often limited by the need for high quality application-specific training data, which is difficult or even impossible to obtain for many dynamic MR applications. Furthermore, even when in-house data can be acquired, sharing is often difficult due to data governance issues. We believe this is a significant barrier to developing DL-based MR reconstructions.

It has previously been shown that static 2D natural images can be used to pre-train DL networks for reconstruction of static 2D MR images(4). The aim of this proof-of-concept study was to train 2D+time DL models from natural videos (e.g moving cars, animals, and people) to reconstruct undersampled dynamic real-time MR data. We used a large open-source database of high-quality natural videos (Inter4K (5)) to train several state-of-the-art DL based reconstructions. These were tested on prospectively acquired real-time cardiac and speech data and compared with DL models trained using dynamic cardiac MR data, as well as CS reconstructions.

# Materials and Methods

This study was approved by the local research ethics committee and written consent was obtained in prospective subjects and retrospective data (ref. 21/EE/0037, 17/LO/1499).

## Experiment Overview

An overview of the study is provided in Figure 1 (details in Supporting Information Figure S1-S3). Three DL architectures were chosen to cover a range of state-of-the-art DL approaches, each paired to a complimentary k-space sampling strategy. These were: 1) an unrolled VarNet (6,7) with a 3D (2D+time) UNet regularizer for Cartesian real-time acquisitions, 2) a 3D (2D+time) multi-coil complex image-based UNet (8,9) for tiny golden-angle radial real-time acquisitions, and 3) a magnitude-only low-latency image-based FastDVDNet (10) network used for spiral real-time acquisitions (HyperSLICE (11)). Each DL method was trained separately on cardiac MR and Inter4K data and evaluated on prospectively acquired real-time cardiac and speech data, along with CS reconstructions.

## Training Dataset Preparation

*Inter4K data:* Simulation of multi-coil k-space data from natural videos is summarized in Figure 2. From the 4K resolution 60 frames-per-second RGB data, two channels were first randomly selected to provide the real and imaginary components with the angle between real and imaginary components scaled to increase phase variations. The resulting complex valued video was cropped (to the size of the real-time MR acquisition) and masked with an ellipse with varying shape and size (to simulate a normal MRI object). Finally, a low frequency random background phase was added to create a more realistic simulated MR object.

The MR acquisition process was then imitated by first generating simulated Gaussian coils with varying intensity, phase, shape, and size. These were applied to the object to obtain multi-coil images, after which independent white Gaussian noise was applied to each coil image. Fourier transformation was then performed to obtain simulated fully-sampled Cartesian dynamic multi-coil k-space. For consistency with the cardiac training dataset, this pipeline was applied to 692 videos from the Inter4K Dataset (Full details are provided in Supporting Information Text S1).

*Cardiac MR data:* All training MRI data were acquired in a single center on a 1.5T system (Aera, Siemens Healthineers, Erlangen, Germany) as previously described (11). The anonymized cardiac MR dataset consisted of 692 ECG-triggered breath-held Cartesian bSSFP CINE multi-coil raw data, acquired in a diverse cardiac patient population in multiple orientations. The raw data was acquired with 2x undersampling (with nominal matrix size of 224x272 and 44 autocalibration lines) and reconstructed with GRAPPA to recover fully-sampled reference Cartesian multi-coil k-spaces.

*Creation of undersampled data:* Processing steps for each architecture were the same for cardiac data and natural videos. VarNet - 2D+time input data was created by compressing multi-coil k-space data to 10 virtual coils (via Singular Value Decomposition) and then undersampling using a variable density Cartesian pattern (17 lines). 3D UNet – Identically compressed multi-coil k-space data was resampled onto 13 tiny golden angle radial spokes and non-uniformly fast Fourier transformed (NUFFT) to produce 2D+time multi-coil complex images. FastDVDNet – Magnitude 2D+time input images were created by resampling k-space data onto 15 variable density spiral interleaves (uniform angle), performing NUFFT and coil combining via root-sum-of-squares. The target outputs were the coil combined (VarNet) or root-sum-of-squares magnitude images (3D UNet and FastDVDNet). An example of natural video zero-filled and DL reconstructions is shown in Supporting Information Video S1.

**Reconstruction and network parameters**

Networks were implemented using TensorFlow/Keras and TensorFlowMRI (12,13) and trained from cardiac data and natural videos using the same parameters. The dataset was split into 519/69/104 videos for training/validation/testing. Networks were trained using a magnitude SSIM loss, an Adam optimizer with a learning rate of $10^{-4}$ for 100, 200 and 200 epochs for the VarNet, 3D UNet and FastDVDNet respectively. In simulations, 24 time-frames were used as input for the VarNet and 3D UNet, while the current and previous four frames were used for the FastDVDNet. The longer prospective time-series were batched for reconstruction. Training and inference were performed on an NVIDIA A6000 GPU.

Code for data preprocessing and training from natural videos and pre-trained networks are available at https://github.com/mrphys/Image_Reconstruction_Inter4k.git.

All reconstructions were tested in simulations using 104 cardiac datasets pre-processed in the same way as the training data and evaluated using Mean Squared Error (MSE), Peak Signal to Noise Ratio (PSNR) and Structural Similarity Index Measure (SSIM). In addition, all datasets were also reconstructed using CS with temporal Total Variation and a regularization factor of $5*10^{-4}$.

**Prospective Assessment**

Ten healthy subjects (30±4 y.o., 73±14 kg) underwent MR on a 1.5T system (same system as training data collected on). Approximately 10 seconds of real-time data were acquired in three separate scan planes: cardiac short axis (SAX), cardiac 4-chamber (4CH), and sagittal head (for speech imaging). In all planes, Cartesian, radial and spiral data were acquired and reconstructed as previously described. Cardiac SAX and 4CH data were acquired using balanced steady-state free precession (bSSFP) during free-breathing, while speech imaging was acquired with spoiled gradient echo (GRE) whilst the subjects performed a simple number counting protocol. More information on the six different prospective real-time acquisitions can be found in Supporting Information Table S1.

*Qualitative image scoring evaluation*: For each scan plane and sampling pattern (Cartesian, radial and spiral), movies of each reconstruction (trained from cardiac data, trained from natural videos and CS) were viewed simultaneously (in a shuffled order) by two clinical experts (V.M., D.K.) and subjectively ranked for overall image quality (1=best, 2= middle, 3=worst, with possibility of tied scores).

*Quantitative image scoring evaluation*: For each scan plane and sampling pattern the reconstructions were compared using: 1) estimated signal-to-noise ratio (SNR) – blood pool mean signal intensity divided by standard deviation of pixels in static region, 2) edge sharpness (ES) – the mean and standard deviation over time of the maximum intensity gradient along normalized profile across the cardiac septum or tongue, and 3) SSIM between images reconstructed using cardiac and natural videos DL (as well as CS) to evaluate similarity of image reconstructions. An example of SNR and ES measurements is provided in Supporting Information Figure S4.

**Statistical Analysis**

As some of the distributions tested were non-normal (Shapiro-Wilk test), a Friedman Chi Square test with post-hoc Nemenyi analysis was used to assess any statistical differences ($p<0.05$) between mean reported metrics.

# Results

*Training and Simulation Results:* Training from natural videos and cardiac images took a similar amount of time, with VarNet, 3D UNet and FastDVDNet taking 32h, 18h and 1h30 to train respectively.

On cardiac MR test data, networks trained with cardiac data outperformed those trained with natural videos, which themselves outperformed CS reconstructions in terms of MSE, PSNR and SSIM (Supporting Information Table S2 and Figure S5). Inference times were similar for both DL models (~10-100x faster than CS, Supporting Information Table S2).

*Prospective Results:* Comparison of the three reconstructions (trained from cardiac data, trained from natural videos and CS) for each sampling pattern (Cartesian, radial, spiral) for representative SAX, 4CH and speech images are shown in Figures 3, 4 and 5 respectively (with corresponding videos in Supporting Information Videos S2, S3, S4).

Qualitative image scoring (Table 1 and 2) for each scan plane and sampling pattern demonstrated that DL networks trained from both cardiac and natural videos ranked significantly higher ($p<0.05$) than CS, except spiral speech (although the same trend is observed). DL reconstructions trained from natural videos ranked similarly to those trained from cardiac data ($p>0.19$) except for Cartesian SAX and 4CH images where they ranked higher ($p=0.04$).

Across all scan planes and sampling patterns, there was no significant difference in SNR between natural video and DL reconstruction from cardiac data, except for Cartesian speech (lower, $p<0.01$) and spiral speech (higher, $p=0.02$). CS had either the lowest SNR or was non-statistically significantly different from DL reconstructions (Table 1 and 2).

For most comparisons, mean ES and the standard deviation of ES over time (STD-ES), between reconstructions were not significantly different (Table 1 and 2, Figures 3-5). A high SSIM index was measured between the DL reconstruction trained from cardiac and natural videos for both prospective SAX and 4CH (>0.9) and speech (>0.85) data.

## Discussion

The main finding of this proof-of-concept study was that it was possible to train DL-based reconstructions for dynamic real-time MR data using open-source natural videos. We used the openly available Inter4K dataset from which we simulated pseudo-MR data, enabling training of a range of state-of the-art complex multi-coil reconstructions (VarNet and 3D UNet) and magnitude only models (FastDVDnet).

We demonstrated that subjective image quality of reconstructions from DL networks trained from natural videos and cardiac data were similar for both real-time cardiac (SAX and 4CH) and speech applications. In most cases this was also reflected in similar quantitative measures of edge sharpness and estimated SNR.

Interestingly, DL networks trained from natural videos had slightly lower image quality than those trained using cardiac data when applied to simulated undersampled cardiac MR test data (when compared to the ground truth). This is unsurprising as DL tends to work better when test inputs (and desired outputs) have a similar distribution to the training data. However, the fact that both DL models performed equally well on prospective data suggests that differences in simulation do not translate to real-world inference. This may be due to distributional shifts between training and prospective data.

Importantly, we have shown that supervised DL models can provide high quality reconstructions even when trained using data that is different from the target output data. Not only have we shown that natural videos can be used to train models for both real-time speech and cardiac applications, but that cardiac MR based DL can be used to successfully reconstruct speech images. This suggests that cardiac MR trained DL models could potentially be used for a wide range of dynamic non-cardiac applications. However, the advantages of using natural videos are open-source availability, easier dissemination, and higher inherent spatial and temporal resolution.

Nevertheless, further study of the significance of the dissimilar image characteristics is required to improve our understanding of both the opportunities and limitations of learning from natural videos(14,15). In addition, improving the quality and generalizability of these models could be explored through scaling up the number of training videos, more general/realistic simulations of coils, noise and object phase and dynamics. Future works should also explore other applications, including the reconstruction of flow data or quantitative maps.

## Conclusion

A proof-of-concept pipeline to learn dynamic MR image reconstruction from publicly available natural videos was applied to a variety of trajectories with different network architectures showing no significant differences or better subjective image quality compared to similar reconstructions trained on true dynamic cardiac MR data.

## Data Availability Statement

The source code for training and testing our framework using natural videos and corresponding pre-trained networks for dynamic MRI reconstruction are provided online (https://github.com/mrphys/Image_Reconstruction_Inter4k.git).

## Acknowledgments

This work used the public dataset Inter4K and open-source framework TensorFlowMRI (v0.22.0).

*Grant Support*: This work was supported by UK Research and Innovation (MR/S032290/1), Heart Research UK (RG2661/17/20) and the British Heart Foundation (NH/18/1/33511, PG/17/6/32797).

# Tables

| Cardiac Table | Reconstruction | SNR (dB) | SSIM | Edge Sharpness (ES) | temporal STD ES | Subjective Image quality ranking |
|---|---|---|---|---|---|---|
| Cartesian | CS | 67.3±16.6*,† | 0.84±0.04 *,† | 0.085±0.021 † | 0.016±0.005 | 2.98±0.16 *,† |
| | Natural Videos | 81.9±11.1 | 0.91±0.02 * | 0.106±0.037 | 0.019±0.006 | 1.20±0.40 * |
| | Cardiac | 92.1±19.0 | 1.00±0.00 | 0.098±0.024 | 0.018±0.008 | 1.75±0.49 |
| Radial | CS | 68.7±10.3*,† | 0.89±0.02 *,† | 0.129±0.032 | 0.017±0.005 | 2.92±0.35 *,† |
| | Natural Videos | 93.8±8.6 | 0.97±0.01 * | 0.127±0.033 | 0.018±0.007 | 1.30±0.51 |
| | Cardiac | 97.0±14.7 | 1.00±0.00 | 0.127±0.029 | 0.017±0.006 | 1.27±0.45 |
| Spiral | CS | 76.6±11.7* | 0.78±0.06 *,† | 0.098±0.028 | 0.015±0.006 † | 2.90±0.30 *,† |
| | Natural Videos | 81.1±8.4 | 0.93±0.01 * | 0.104±0.034 | 0.020±0.007 | 1.55±0.59 |
| | Cardiac | 84.0±12.6 | 1.00±0.00 | 0.098±0.024 | 0.017±0.005 | 1.50±0.59 |

Table 1: Mean and standard deviation values for prospective cardiac images (SAX, 4CH, N=20). From left to right: SNR (in decibels -dB), SSIM, Edge Sharpness (mean over 1.5s), temporal standard deviation of edge sharpness through time (std over 1.5s) and subjective image quality ranking.

*,† Statistically significantly different from cardiac and natural videos respectively ($p<0.05$).

| Speech Table | Reconstruction | SNR (dB) | SSIM | Edge Sharpness (ES) | temporal STD ES | Subjective Image quality ranking |
|---|---|---|---|---|---|---|
| Cartesian | CS | 53.0±5.1* | 0.77±0.03 * | 0.157±0.023 | 0.046±0.017 | 2.95±0.22 *,† |
| | Natural Videos | 53.9±4.3* | 0.85±0.03 | 0.169±0.024 | 0.049±0.018 | 1.15±0.36 |
| | Cardiac | 65.8±4.7 | 1.00±0.00 | 0.162±0.031 | 0.046±0.016 | 1.7±0.46 |
| Radial | CS | 41.0±3.3*,† | 0.67±0.05 * | 0.184±0.032 | 0.045±0.014 | 3.00±0.0 *,† |
| | Natural Videos | 83.1±2.4 | 0.96±0.01 | 0.183±0.036 | 0.051±0.018 | 1.15±0.36 |
| | Cardiac | 89.3±5.5 | 1.00±0.00 | 0.175±0.029 | 0.045±0.019 | 1.25±0.43 |
| Spiral | CS | 68.8±6.9† | 0.50±0.06 * | 0.114±0.014 | 0.033±0.013 | 2.30±0.84 |
| | Natural Videos | 73.7±7.4* | 0.93±0.01 | 0.119±0.018 | 0.034±0.013 | 1.70±0.78 |
| | Cardiac | 66.5±15.0 | 1.00±0.00 | 0.117±0.02 | 0.038±0.018 | 2.10±0.77 |

Table 2: Mean and standard deviation values for prospective speech images (N=10). From left to right: SNR (in decibels -dB), SSIM, Edge Sharpness (mean over 1.5s), temporal standard deviation of edge sharpness through time (std over 1.5s) and subjective image quality ranking.

*,† Statistically significantly different from cardiac and natural videos respectively ($p<0.05$).

# Figures

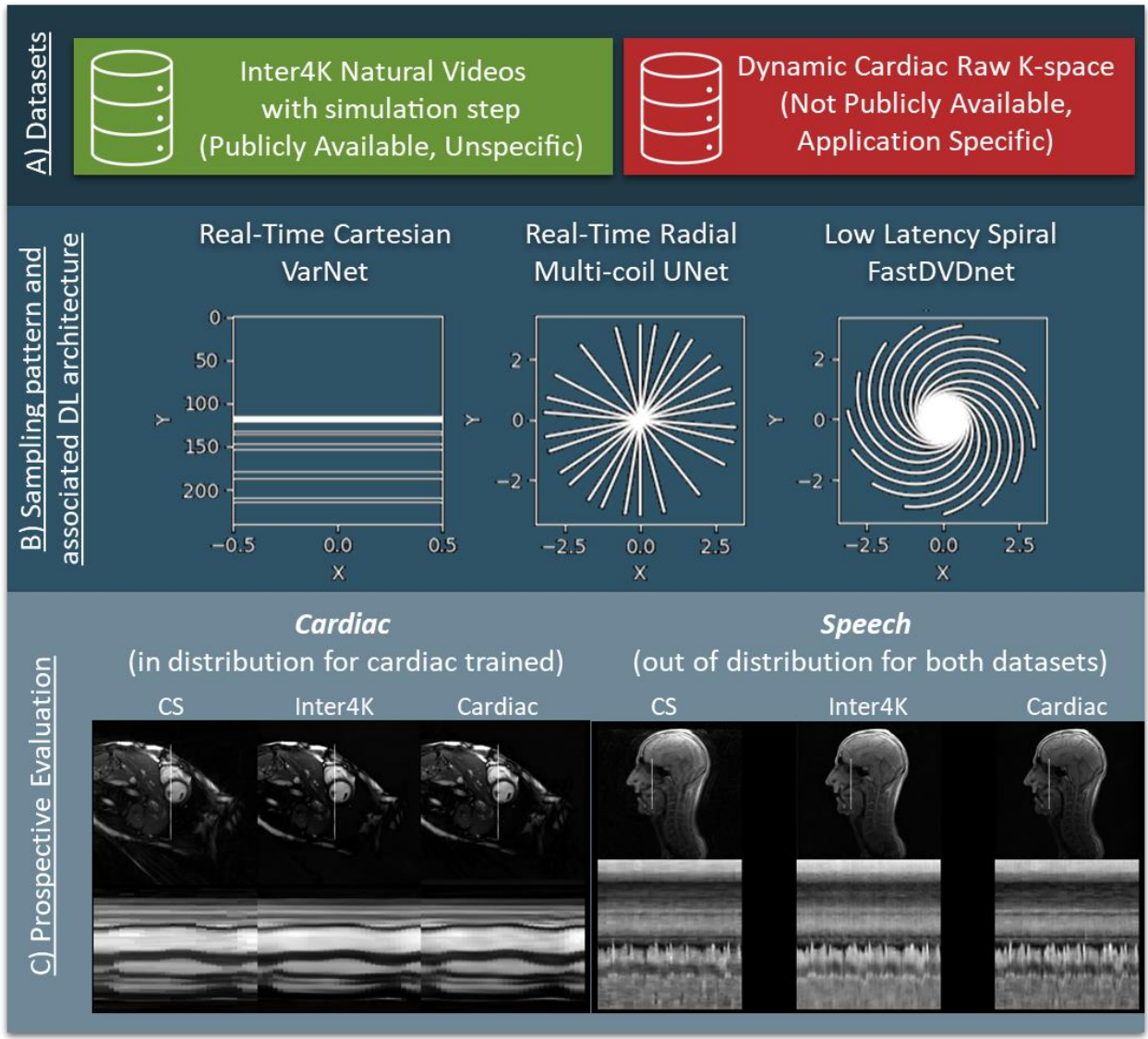

Figure 1. Experiment Overview. Training reconstructions from the natural video Dataset (Inter4K) was compared to training from a true cardiac MR dataset for three different sampling patterns and associated supervised DL reconstructions of real-time data: Cartesian acquisition with VarNet 2D+time reconstruction, radial acquisition with multi-coil 3D UNet reconstruction and spiral acquisition with FastDVDNet reconstruction. These methods were tested against Compressed Sensing (CS) reconstructions on two prospective applications: cardiac bSSFP in SAX and 4CH views (similar to cardiac training data) and Speech GRE (out of distribution for both cardiac and natural videos).

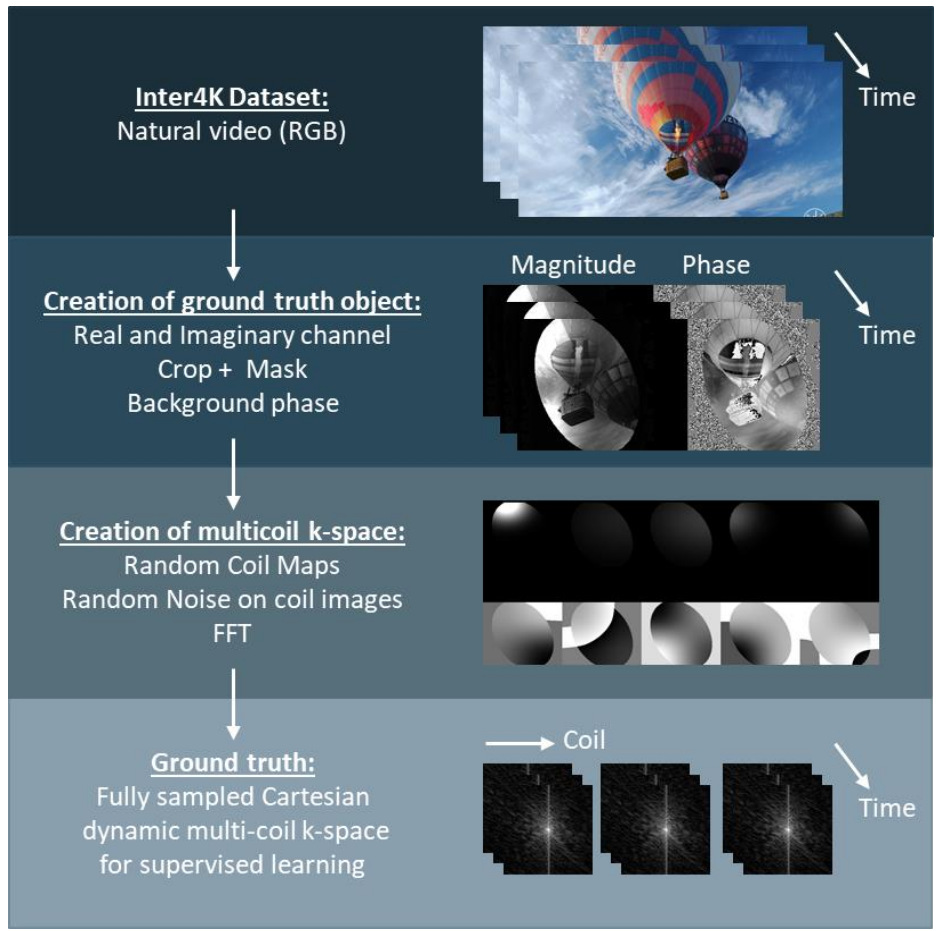

Figure 2. Pipeline for generating fully sampled Cartesian dynamic multi-coil k-space from natural videos. Step-by-step: 1) Creation of ground truth object: 2 of the 3 channel RGB video are randomly selected to create the real and imaginary parts and the phase between the two is scaled. Images are cropped and masked with an elliptical mask. Low frequency background phase is added. 2) Creation of ground truth multi-coil k-space: Random coil maps are generated and applied to the object, random noise is added on each coil image separately and finally a fast Fourier transform is applied to obtain the k-space data. 3) Finally, the data can be used as a regular input to any supervised reconstruction method. More details in Supporting Information Text S1.

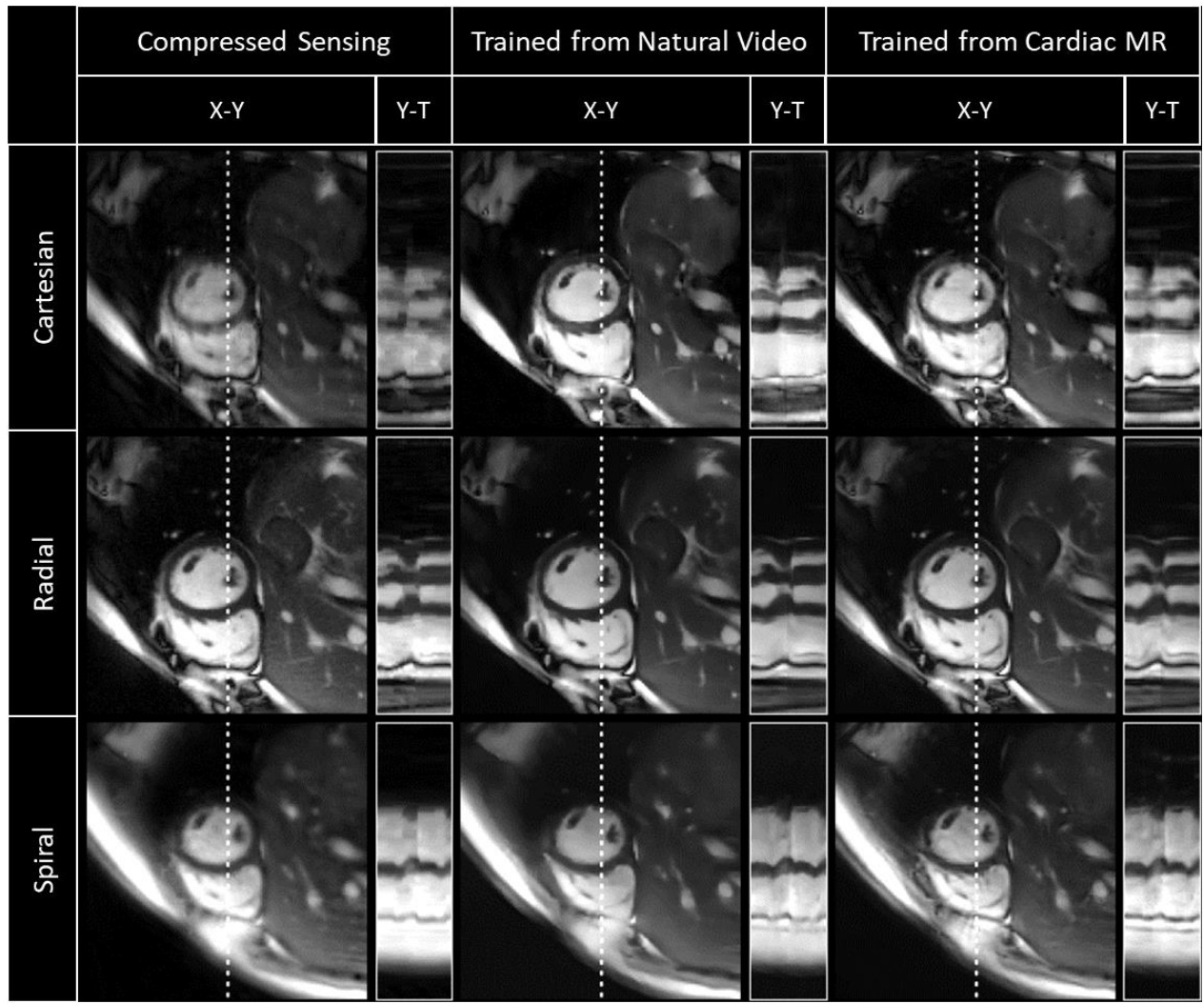

**Figure 3. Qualitative comparison of a cardiac short axis dataset showing cropped image and Y-T profiles as indicated by white dotted line.** From Top to Bottom: Real-time Cartesian, radial and spiral prospective acquisitions. From left to right: Compressed Sensing, natural video trained and Cardiac trained reconstructions. Deep learning architectures were VarNets, multi-coil 3D UNet, and low latency FastDVDNet for Cartesian, radial and spiral respectively. Corresponding video can be found in Supporting Information Video S2.

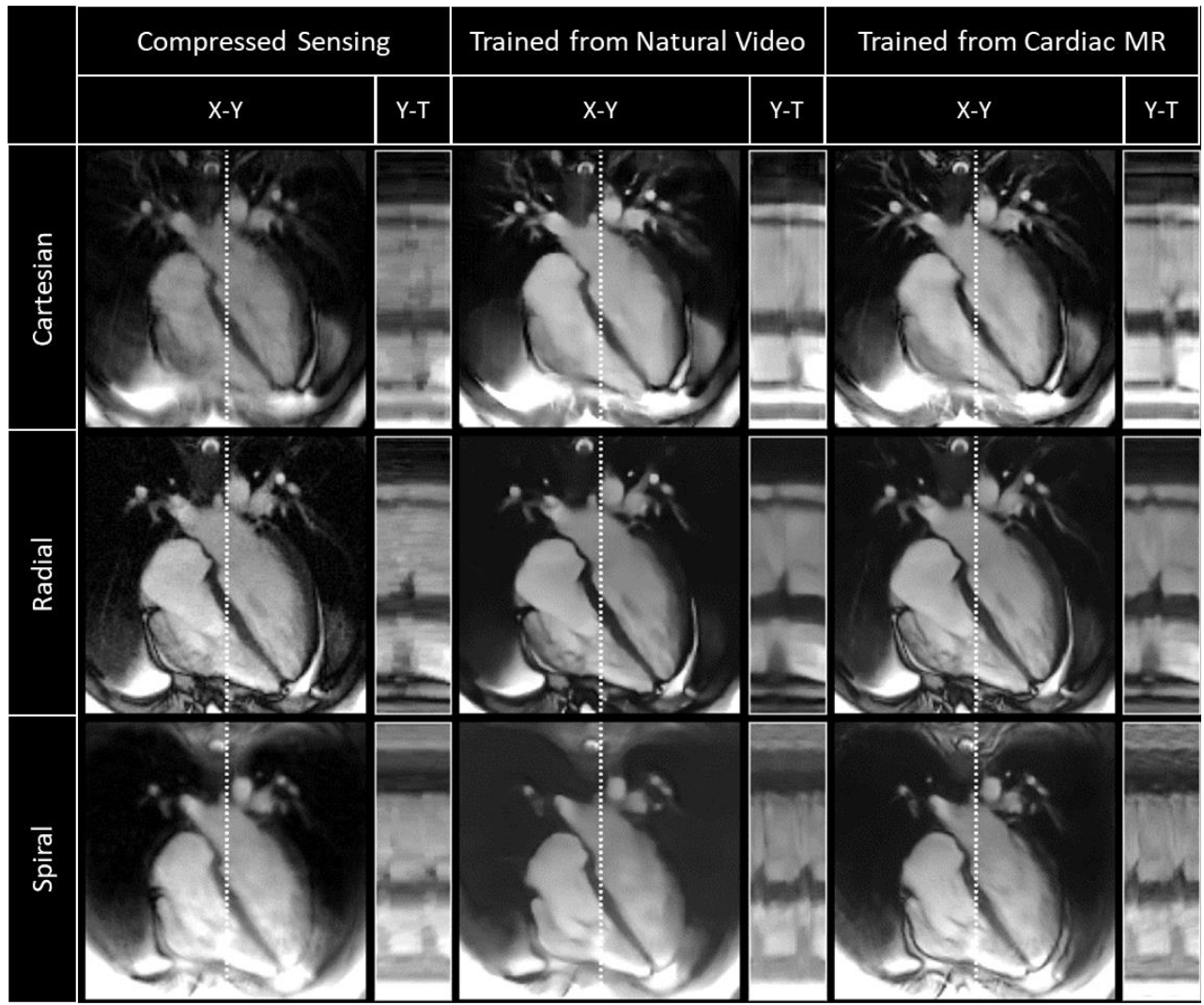

**Figure 4. Qualitative comparison of a cardiac four chamber dataset showing cropped image and Y-T profiles as indicated by white dotted line.** From Top to Bottom: Real-time Cartesian, radial and spiral prospective acquisitions. From left to right: Compressed Sensing, natural video trained and cardiac trained reconstructions. Deep learning architectures were VarNets, multi-coil 3D UNet, and low latency FastDVDNet for Cartesian, radial and spiral respectively. Corresponding video can be found in Supporting Information Video S3.

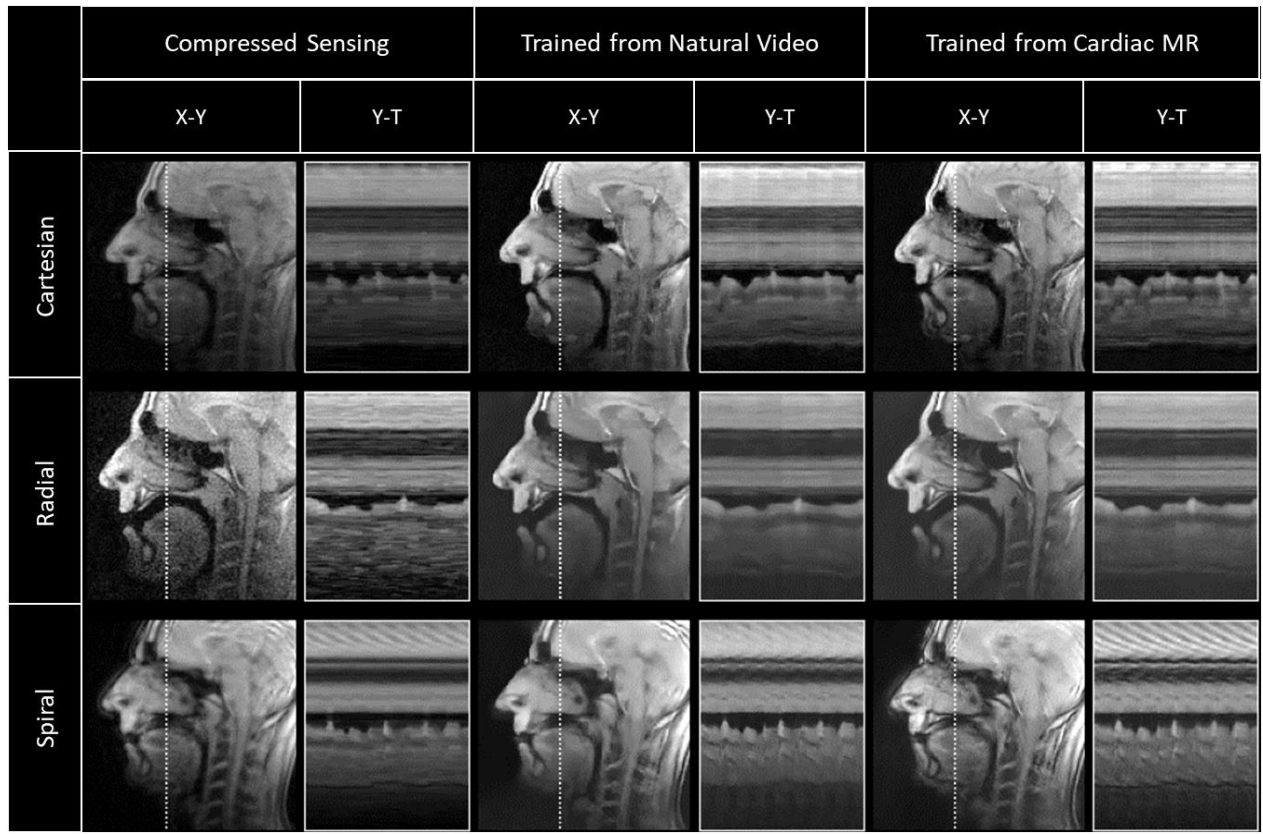

**Figure 5. Qualitative comparison of a speech dataset showing cropped image and Y-T profiles as indicated by white dotted line.** From Top to Bottom: Real-time Cartesian, radial and spiral prospective acquisitions. From left to right: Compressed Sensing, natural video trained and cardiac trained reconstructions. Deep learning architectures were VarNets, multi-coil 3D UNet, and low latency FastDVDNet for Cartesian, radial and spiral respectively. Corresponding video can be found in Supporting Information Video S4.

# Supplemental Material

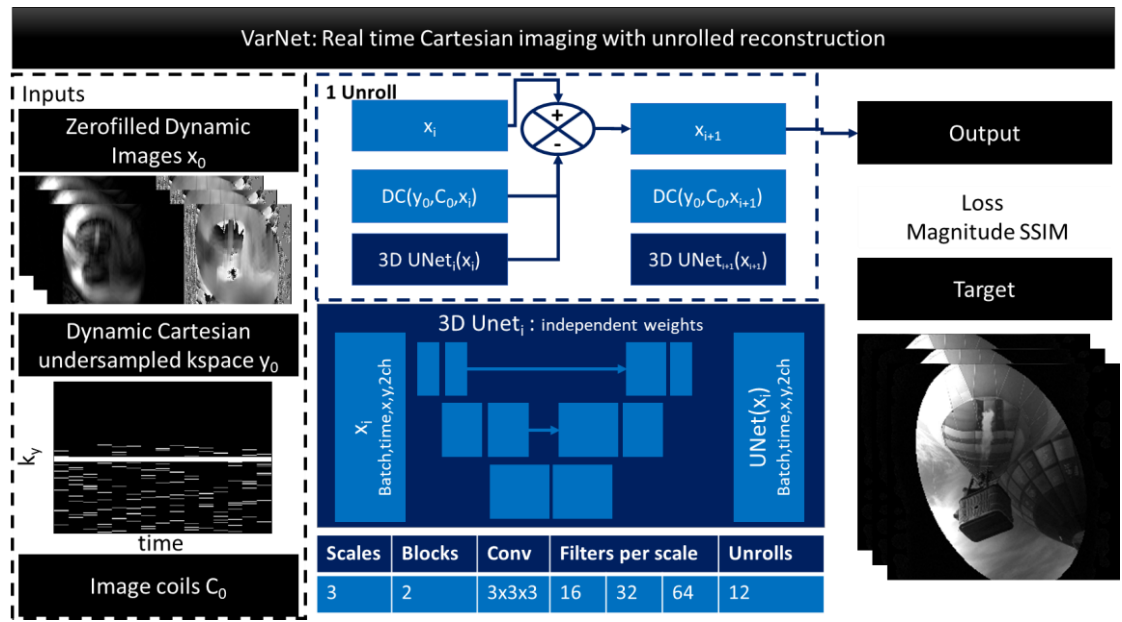

**Supporting Information Figure S1. Summarized method for the Cartesian acquisition with VarNet reconstruction network.** The inputs consist of 24 zero-filled coil combined consecutive images as initialization $x_0$, corresponding complex multi-coil k-space $y_0$ with 17 lines per timepoint (including 8 center lines + 9 randomly non-repeating lines sampled in bottom 60% of k-space), and coil sensitivities estimated from the time combined data. The VarNet architecture was extended to 2D+time by applying a 3D UNet for regularization. Parameters are included in the bottom table. The network was trained using an ADAM optimizer for 100 epochs using a Magnitude SSIM loss.

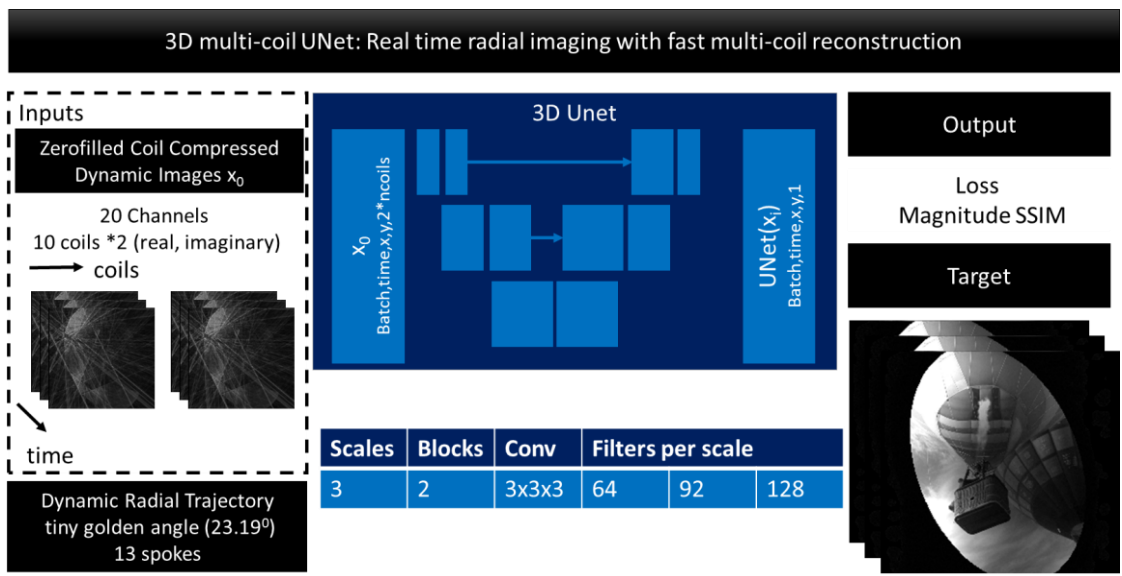

**Supporting Information Figure S2. Summarized method for the Radial acquisition with multi-coil 3D UNet reconstruction.** The input consists of 24 gridded complex consecutive images coil compressed to 10 coils. Each frame was acquired using 13 spokes incremented by the tiny golden angle in k-space (23.19°). UNet parameters are included in the bottom table. The network was trained using an ADAM optimizer for 200 epochs using a Magnitude SSIM loss.

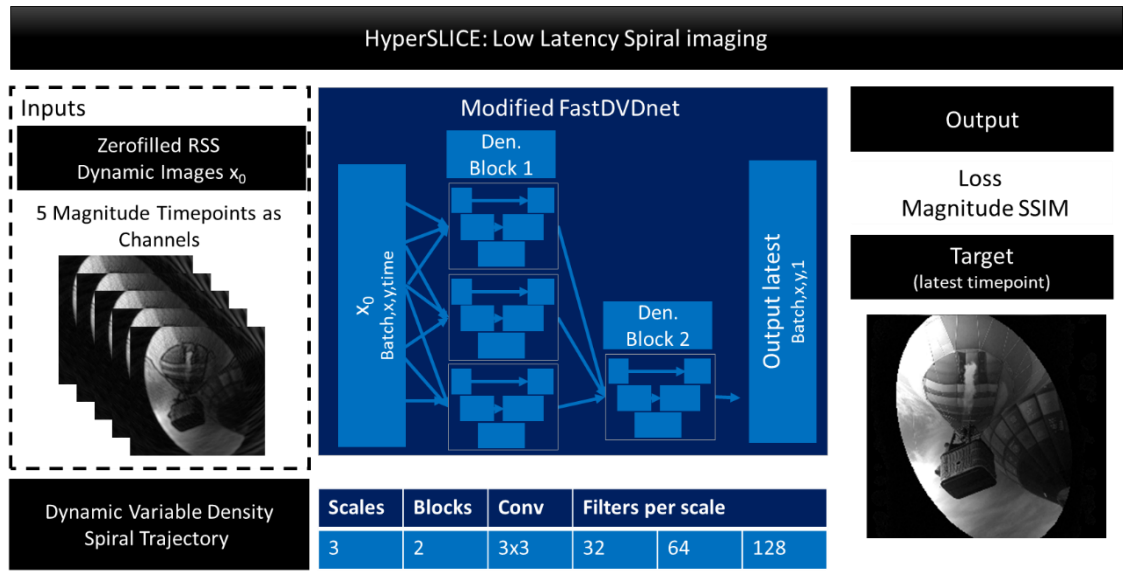

**Supporting Information Figure S3. Summarized method for the spiral acquisition with low latency FastDVnet reconstruction.** The input consists of the five latest gridded root-sum-of-squares magnitude images. Each frame was acquired using a variable density spiral trajectory in k-space. Denoising block parameters are included in the bottom table. The network was trained to reconstruct the latest timepoint using an ADAM optimizer for 200 epochs using a Magnitude SSIM loss. More details on the acquisition and reconstruction can be found in (11).

**Supporting Information Text S1. Pipeline for generating fully sampled Cartesian dynamic multi-coil k-space from natural videos.**

1) **Preprocessing:**
    - First 50 consecutive frames are selected.
    - Downsampling is performed frame per frame via 2D bilinear interpolation to a fixed image size.
2) **Creation of ground truth object:**
    - Two of the three channel RGB video are randomly selected to create the real and imaginary parts.
    - The phase between the two components is scaled by a factor of 4 (empirically chosen).
    - Images are cropped to final image size (depends on acquisition) and masked with a randomly rotated elliptical mask with long and short axis in range [1.0:1.4] and [0.64:0.96] times the image width respectively.
    - Finally low frequency background phase is created by upscaling a 6x6 random matrix using bicubic interpolation and added to the image phase.
3) **Creation of ground truth multi-coil k-space:**
    - 30 random coil maps are generated by creating 2D Gaussians with random maximum intensity [0.1:1], random shape with standard deviation for x and y axes in the range [0.5:0.16] times the image size, random center location (avoiding 1/5[th] of image center), and random offset phase and background phase. Finally coil maps are normalized by the root-sum-of-squares of all the generated coil maps.
    - Coil images are obtained by multiplying each coil map to the image series.
    - Uniform random noise unique to each coil and timepoint images is added to reach an average target SNR in the range of [12:22].
    - Finally a fast Fourier transform is applied to obtain the k-space data.
4) **The data can be used as a regular input to any supervised reconstruction method.**

| Trajectory | Sampling | Application | Type | Temporal resolution (ms) | Flip Angle (°) | TR (ms) | Spatial resolution | Field of view (mmxmm) |
|---|---|---|---|---|---|---|---|---|
| Cartesian | 17 lines including<br>- 8 in center<br>- 9 uniform random samples in half Fourier (no resampling) | Cardiac | bSSFP | 47 | 70 | 2.8 | 1.7x1.7 | 400x400 |
| | | Speech | GRE | 58 | 15 | 3.4 | 1.7x1.7 | 400x400 |
| Radial | 13 spokes<br>Tiny golden angle increment (23.8°) | Cardiac | bSSFP | 38 | 70 | 3.0 | 1.6x1.6 | 400x400 |
| | | Speech | GRE | 45 | 15 | 3.4 | 1.6x1.6 | 400x400 |
| Spiral | 15 variable density arms<br>Variable Acceleration<br>- R=[1.1 inner, 15 outer]<br>- Inner radius=15% of k-space<br>- Outer radius=56% of k-space | Cardiac | bSSFP | 55 | 70 | 3.7 | 1.7x1.7 | 400x400 |
| | | Speech | GRE | 79 | 15 | 5.3 | 1.7x1.7 | 400x400 |

**Supporting Information Table S1. Acquisition details for the six different prospective real-time acquisitions.** Cardiac and speech had the same trajectories and therefore were reconstructed using the same network however due to the change from bSSFP to GRE the temporal resolution and flip angle were different.

A) **Cartesian:** 17 lines are acquired per timepoint including the 8 center lines and 9 randomly selected samples in bottom 60% of Fourier space. The samples cannot be selected twice within any consecutive 15 frames leading to an almost fully sampled cartesian k-space after 15 frames.

B) **Radial:** 13 radial spokes are sampled with a constant tiny golden angle increment of 23.8°.

C) **Spiral:** 15 variable density spiral arms are sampled for each cardiac phase. The final acceleration rate for a single cardiac phase is of 1.1 in the inner 15% of k-space and 15 above the 56% of k-space radius. The same spiral arm is acquired with linear rotation between spiral arms of the same timepoint (by 1/15th of a full rotation to cover full k-space for one cardiac phase) and is also shifted linearly between timepoints (by 1/12th). The trajectory repeats every 12 phases.

| Acquisition | Reconstruction | MSE *$10^4$ | PSNR | SSIM | Inference time |
|---|---|---|---|---|---|
| Cartesian | CS Temporal TV | 15.72±8.49 | 28.68±2.39 | 0.8±0.07 | 8.2s per batch |
| | VarNet Natural Videos | 8.57±4.67 | 31.18±2.04 | 0.82±0.06 | 0.9s per batch |
| | VarNet Cardiac | 3.61±2.3 | 35.16±2.48 | 0.9±0.05 | 0.9s per batch |
| Radial | CS Temporal TV | 6.96±5.58 | 32.63±3.06 | 0.8±0.07 | 10.04s per batch |
| | 3D MC UNet Natural Videos | 3.67±2.16 | 35.06±2.45 | 0.88±0.07 | 0.11s per batch |
| | 3D MC UNet Cardiac | 2.01±1.3 | 37.8±2.63 | 0.93±0.04 | 0.11s per batch |
| Spiral | CS Temporal TV | 14.91±14.49 | 29.38±3.0 | 0.69±0.08 | 1395 ms/frame (33.4s per batch) |
| | FastDVDnet Inter4k | 6.51±3.04 | 32.31±1.95 | 0.86±0.05 | 22 ms/frame |
| | FastDVDnet Cardiac | 4.15±2.2 | 34.39±2.26 | 0.9±0.03 | 22 ms/frame |

**Supporting Information Table S2.** Simulation mean and standard deviation metric values from test set (N=104). From left to right: MSE *$10^4$, PSNR, SSIM and inference time.

All results for MSE, PSNR and SSIM were statistically significantly different (p<0.05) between the 3 reconstructions.

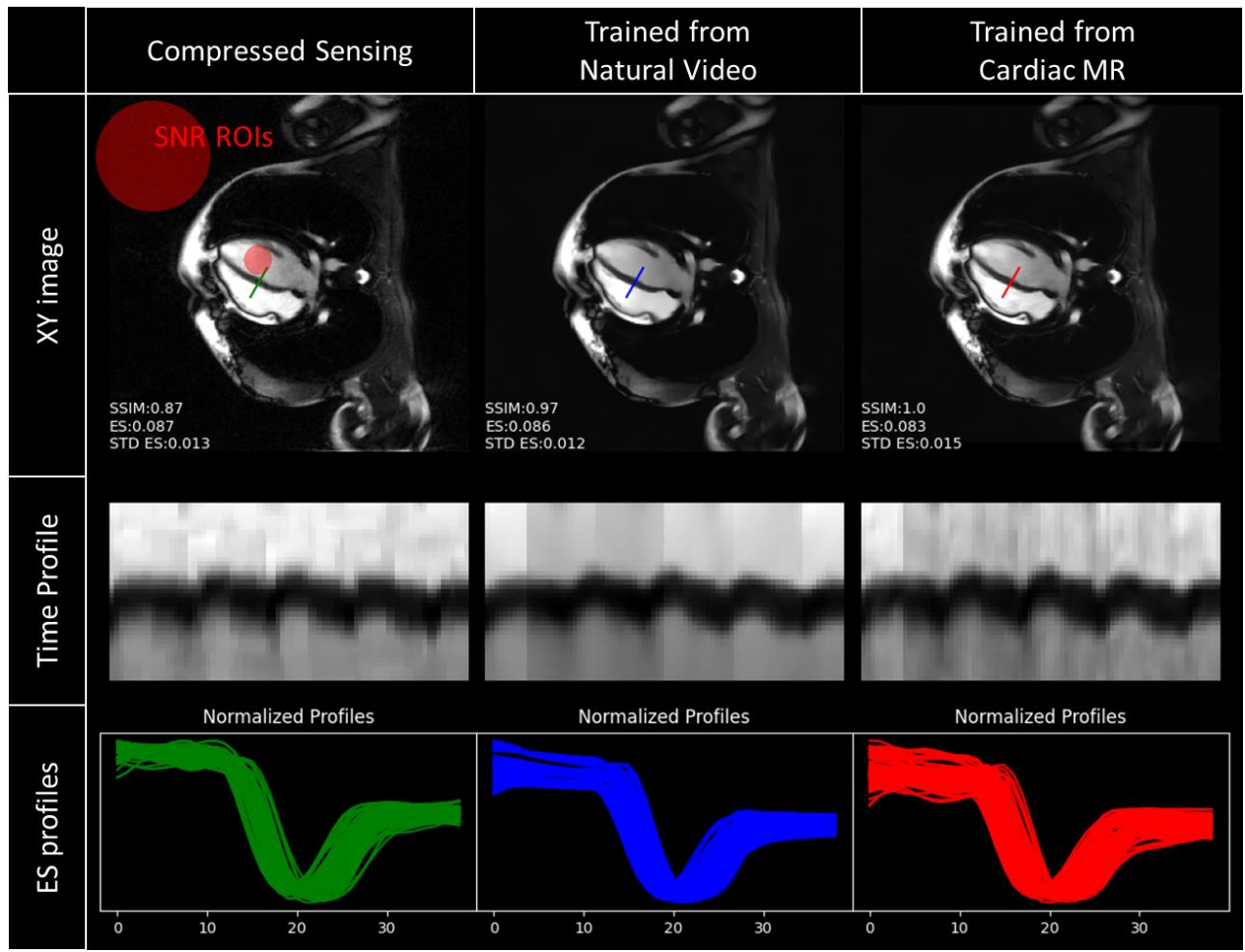

**Supporting Information Figure S4.** Example of quantitative measurements performed in one radial prospective 4chamber dataset for the three reconstructions considered: Compressed Sensing with temporal TV regularization, 3D UNet trained with cardiac data and 3D UNet trained with natural videos. Representative examples of measurements of SNR (ROIs) and edge sharpness mean (ES) and temporal standard deviation (STD ES) using the maximum gradient of the normalized profiles across the septum for each timepoint.

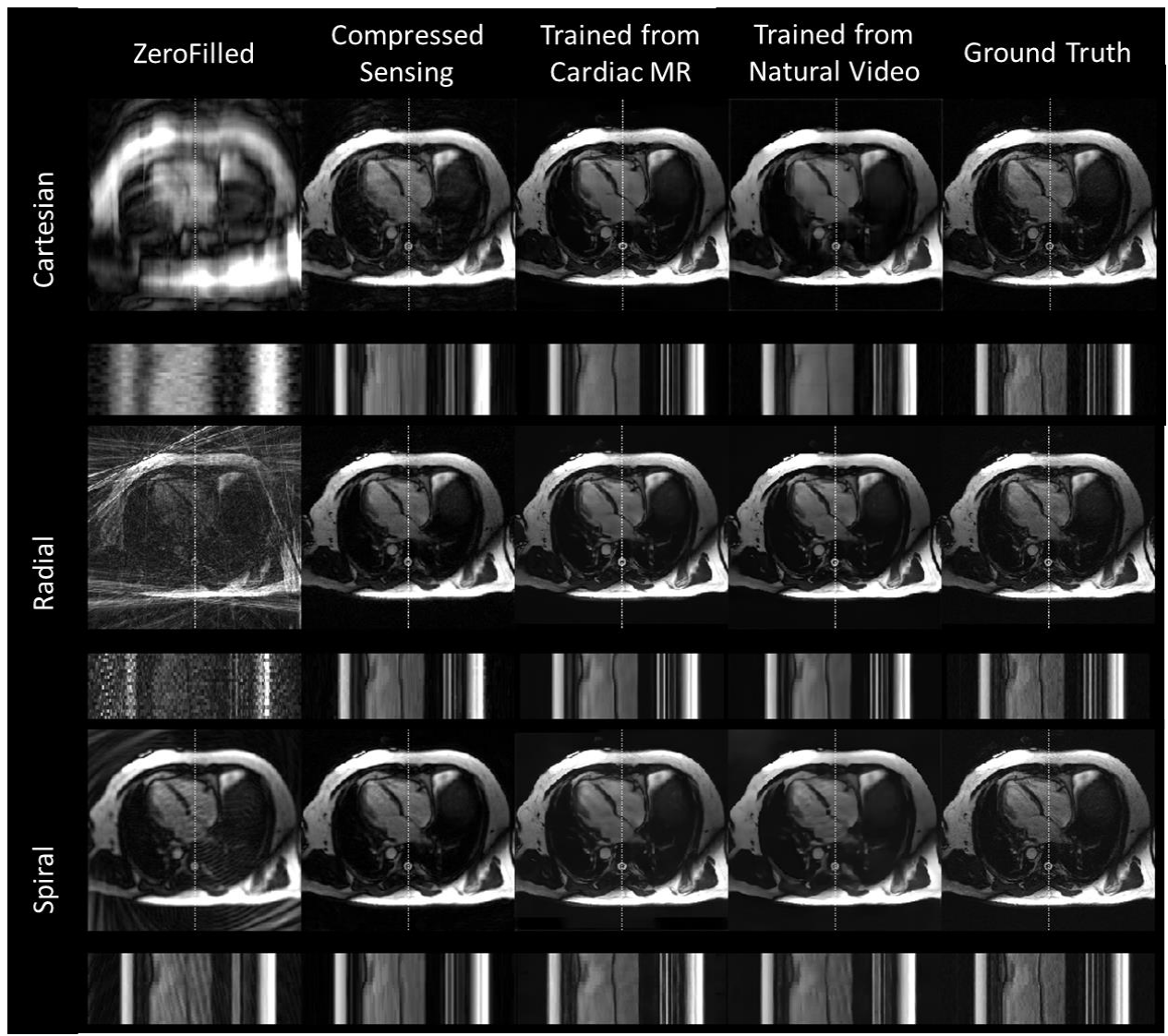

**Supporting Information Figure S5.** Qualitative Results in one test set comparing Zero-filled, Compressed Sensing, cardiac trained and natural video trained reconstructions to the Ground Truth for the three methods considered (Cartesian VarNet, Radial 3D UNet, Spiral FastDVDNet).

**Supporting Information Video S1. Overview of Inter4K video and reconstructions from a test sample.** Top: Cropped RGB, undersampled Cartesian zero-filled reconstruction, undersampled radial zero-filled reconstruction, undersampled spiral zero-filled reconstruction. Bottom: Target magnitude image, VarNet reconstruction, multi-coil 3D Unet reconstruction, FastDVDnet reconstruction.

**Supporting Information Video S2. Video comparison of a short axis cardiac dataset.** From Top to Bottom: Real-time Cartesian, radial and spiral prospective acquisitions. From left to right: Compressed Sensing, natural video trained and cardiac trained reconstructions. Machine learning reconstructions were VarNets, multi-coil 3D UNet, and low latency FastDVDNet for Cartesian, radial and spiral respectively.

**Supporting Information Video S3. Video comparison of a four chambers cardiac dataset.** From Top to Bottom: Real-time Cartesian, radial and spiral prospective acquisitions. From left to right: Compressed Sensing, natural video trained and cardiac trained reconstructions. Machine learning reconstructions were VarNets, multi-coil 3D UNet, and low latency FastDVDNet for Cartesian, radial and spiral respectively.

**Supporting Information Video S4. Video comparison of a speech dataset.** From Top to Bottom: Real-time Cartesian, radial and spiral prospective acquisitions. From left to right: Compressed Sensing, natural video trained and cardiac trained reconstructions. Machine learning reconstructions were VarNets, multi-coil 3D UNet, and low latency FastDVDNet for Cartesian, radial and spiral respectively.